\documentclass[epj]{svjour}

\usepackage{graphicx}
\usepackage{fancyhdr}

\def\mytitle{Unification with and without Supersymmetry:Adjoint SU(5)} 
\def\myauthors{Pavel Fileviez P\'erez}  
\def\mytype{Parallel Session}
\def\mysession{Alternatives}
\def\mytitle{Unification with and without Supersymmetry: Adjoint SU(5)} %Put your title here!
\def\myauthors{Pavel Fileviez P\'erez}    %Put your name here!
\def\mytype{Contributed Talk}    
\def\mysession{Alternatives}
\begin{document}
%%%%%%%%%%%%%%%%%%%%%%%%%%%%%%%%%%%%%%%%%%%%%%%%%%%%%%%%%%%%%%%%%%

\title{Unification with and without Supersymmetry: Adjoint SU(5)}
\author{Pavel Fileviez P\'erez \inst{}
\thanks{\small Talk given at the 15th International Conference on Supersymmetry 
and the Unification of Fundamental Interactions, Karlsruhe, Germany, July 26 - August 1, 2007.
\emph{Email:} fileviez@physics.wisc.edu}}
\institute{University of Wisconsin-Madison, Department of Physics, 
1150 University Ave. Madison, WI 53706, USA
\and 
CFTP, \ Departamento de F{\'\i}sica,\ Instituto Superior T\'ecnico,
\ 1049-001 Lisboa, Portugal
}
\date{}
%%%%%%%%%%%%%%%%%%%%%%%%%%%%%%%%%%%%%%%%%%%%%%%%%%%%%%%%%%%%%%%%%%
\abstract{
I present two new renormalizable grand unified theories where 
the neutrino masses are generated through the type I and type III 
seesaw mechanisms. These theories can be considered as the 
simplest (SUSY) renormalizable grand unified theories 
based on the SU(5) gauge symmetry. Several phenomenological and 
cosmological aspects of these proposals are discussed.
\PACS{
      {12.10.Dm}{Unified theories and models of strong and electroweak interactions}  
     } % end of PACS codes
} 
%end of abstract
%
\maketitle

%%%%%%%%%%%%%%%%%%%%%%%%%%%%%%%%%%%%%%%%%%%%%%%%%%%%%%%%%%%%%%%%
\section{Introduction}
%%%%%%%%%%%%%%%%%%%%%%%%%%%%%%%%%%%%%%%%%%%%%%%%%%%%%%%%%%%%%%%%
\label{intro}
The unification of the electromagnetic, weak and strong 
interactions is one of the main motivations for 
physics beyond the Standard Model. The candidates which 
describe physics at the unification scale, $M_{GUT} \approx 10^{14-16}$ GeV, 
are called Grand Unified Theories (GUTs). The simplest theories 
based on $SU(5)$ or $SO(10)$ have been proposed a long time ago. 
However, we still do not know which is the best candidate 
for grand unification and all possibilities 
to test or rule out the simplest theories. Usually 
the most promising way to test grand unified 
theories is through proton decay~\cite{review}. 

Grand unified theories predict the unification of 
the Standard Model interactions at the GUT scale, 
the quantization of the electric charge, the value 
of $\sin^2 \theta_W (M_{GUT}) = 3/8$ at the GUT scale, 
the decay of the proton and the existence of leptoquarks. 
I will focus on the grand unified theories based on $SU(5)$ 
since those are the simplest candidates for grand unification. 

The simplest grand unified theory was proposed in 
reference~\cite{GG}. This theory is based on the $SU(5)$ gauge 
symmetry and one Standard Model family is partially unified 
in the anti-fundamental ${\bf \overline{5}}$ and antisymmetric 
${\bf 10}$ representations. The Higgs sector is composed of 
${\bf {{24}}_H}$ and ${\bf {{5}}_H}$, the GUT symmetry is broken 
down to the Standard Model by the vacuum expectation value of the 
Higgs singlet field in ${\bf {24}_H}$, and the Standard Model 
Higgs resides in ${\bf {{5}}_H}$. Unfortunately, this theory is 
ruled out since in this case the unification of gauge couplings 
is in disagreement with the values of $\alpha_s (M_Z)$, 
$\sin \theta_W (M_Z)$ and $\alpha_{em}(M_Z)$. Therefore, 
we have to look for realistic GUTs based on SU(5).  

The minimal supersymmetric version of the Georgi-Glashow model 
was discussed for the first time in reference~\cite{SUSYSU(5)}. In this 
case one generation of matter of the Minimal Supersymmetric 
Standard Model (MSSM) is unified in two chiral superfields 
$\mathbf{\hat{\bar 5}} = (\hat{D}^C,\hat{L})$ and 
$\mathbf{\hat{10}} = (\hat{U}^C,\hat{Q},\hat{E}^C)$, 
while the Higgs sector is composed of $\mathbf{\hat 5_H}=(\hat{T}, 
\hat{H}_1)$, $\mathbf{\hat{\bar 5}_H}=(\hat{\overline{T}},\hat{\overline{H}}_1)$, 
and $\mathbf{\hat{24}_H}$. The renormalizable version 
of this theory is ruled out since the relation between $Y_E$ 
and $Y_D$, $Y_E=Y_D^T$, is in disagreement with the experimental 
values of the fermion masses at the low scale and the neutrinos 
are massless if the so-called R-parity is conserved~\footnote{ 
\small See reference~\cite{SUSY-IPG} for the most general constraints 
coming from unification and~\cite{BPG} for all possible dimension 
five contributions to the decay of the proton in this context.}. 

As we have mentioned above, the main and common problems 
of the Georgi-Glashow model~\cite{GG} and minimal 
renormalizable SUSY $SU(5)$~\cite{SUSYSU(5)} are the relation 
between $Y_E$ and $Y_D$: $Y_E=Y_D^T$, and the absence of 
neutrino masses. In the context of $SU(5)$ models 
there are two possible ways to obtain a consistent relation 
between the masses for down quarks and charged leptons: 
i) one introduces an extra Higgs in the $45$ representation~\cite{GJ} 
or ii) one takes into account the effect of higher-dimensional 
operators~\cite{Ellis}. Notice that in the first possibility we 
can write a consistent model and keep renormalizability, while 
in the second case it is difficult to know which is the 
effect of higher-dimensional operators in all 
sectors of the theory. Now, in order to generate neutrino masses 
at tree level in this context one can study the implementation 
of the Type I~\cite{TypeI}, Type II~\cite{TypeII} or 
Type III~\cite{TypeIII} seesaw mechanisms. Different combinations 
of the needed mechanisms to solve the problem of fermion masses 
give us the possibility to write down several realistic 
grand unified theories which could be tested. 
Let us discuss the different possibilities:
\begin{itemize}

\item \textit{Renormalizable Non-SUSY SU(5) models}. 
The Higgs sector of those models will be composed at least of 
${\bf 5_H}$, ${\bf 24_H}$ and ${\bf 45_H}$. Now, we can have three 
different scenarios:
\\
\begin{itemize}

\item \textit{ Model with Type I seesaw}: 
In this case at least two extra fermionic singlets are 
included in the theory. See reference~\cite{Ilja-Pavel-45} 
for a recent study.\\   

\item \textit{ Model with Type II seesaw}: In this scenario the 
Higgs sector is extended adding a Higgs in the ${\bf 15}$ representation. 
See reference~\cite{Ilja-Pavel-45}.\\

\item \textit{  Model with Type III seesaw}: In order to generate 
neutrino masses a fermionic multiplet in the adjoint 
representation is added. In this talk I will discuss 
in detail this model. We refer to this model as 
``Renormalizable Adjoint $SU(5)$''~\cite{paper1}.  \\
\end{itemize}

\item \textit{Non-Renormalizable (NR) SU(5) models}. 
In this case the Higgs sector is composed of ${\bf 5_H}$ and ${\bf 24_H}$, and 
the effect of higher-dimensional operators is taking into account. \\

\begin{itemize}

\item  \textit{ NR Model with Type I seesaw}: This model 
is ruled out by unification as the original model proposed by Georgi and Glashow.\\ 

\item \textit{ NR Model with Type II seesaw model}: This 
theory has been proposed in reference~\cite{Ilja-Pavel} and 
studied in detail in~\cite{IPR,IPG,P}. In this case it is possible to achieve 
unification in agreement with the experiments, the model 
predicts for the first time the existence of light scalar 
leptoquarks and the upper bound on the total proton decay 
lifetime~\cite{upper} is $\tau_p^{upper} \leq 2 \times 10^{36}$ years. 
Therefore, we could have the possibility to test the idea of 
grand unification at future colliders since the light 
scalar leptoquarks, $\Phi_b=(3,2,1/6)$, could be produced 
at the Large Hadron Collider (LHC) or at $e^+ e^-$ colliders. \\

\item \textit{ NR Model with Type III seesaw}: This model has been 
proposed in reference~\cite{Borut-Goran}. See also 
reference~\cite{Ilja-Pavel-24F} for the predictions coming 
from the unification of gauge couplings in this context. \\

\end{itemize}

Now, let us discuss the different supersymmetric models based on SU(5).
\\
\item \textit{Renormalizable SUSY SU(5) models}. In these models 
the Higgs sector is composed of ${\bf \hat{5}_H}$, ${\bf \hat{\overline{5}}_H}$, ${\bf \hat{24}_H}$, 
${\bf \hat{45}_H}$ and ${\bf \hat{\overline{45}}_H}$, and one can have different scenarios:\\

\begin{itemize}

\item \textit{ SUSY Model with Type I seesaw}: In this case we have to 
add at least two singlets superfields in order to generate 
neutrino masses. \\

\item \textit{ SUSY Model with Type II seesaw}: The Higgs sector is 
extended by adding two chiral superfields, ${\bf \hat{15}_H}$ and 
${\bf \hat{\overline{15}}_H}$. See reference~\cite{Rossi} for the study 
of this scenario. \\

\item \textit{ SUSY Model with Type III seesaw}: This model has been 
proposed in reference~\cite{paper2} and will be discussed 
in the next sections. We refer to this model as 
``Supersymmetric Adjoint $SU(5)$''~\cite{paper2}.\\
 
\end{itemize}
\end{itemize}
We can also write different Non-Renormalizable SUSY $SU(5)$ models 
where the effect of higher-dimensional operators is considered and the 
neutrino masses is generated through the different seesaw mechanisms. 
We do not list here all possibilities since in this case we just replace 
the Higgs chiral superfields ${\bf \hat{45}_H}$ and ${\bf \hat{\overline{45}}_H}$ by 
the higher dimensional operators. See for example~\cite{Nath96}. 

After this short review I will discuss two new grand unified theories 
based on the $SU(5)$ gauge symmetry. The first theory has been 
proposed in reference~\cite{paper1} and the fermion masses are 
generated with the minimal set of Higgs bosons, ${\bf \mathbf{5_H}}$ 
and ${\bf \mathbf{45_H}}$. The neutrino masses are generated through 
the type I~\cite{TypeI} and type III~\cite{TypeIII} seesaw mechanisms 
using the fermionic ${\bf \mathbf{24}}$ representation. In the second 
section we will discuss the SUSY version of this model and we will show 
that these theories can be considered as the simplest (SUSY) renormalizable 
grand unified theories based on $SU(5)$.

%%%%%%%%%%%%%%%%%%%%%%%%%%%%%%%%%%%%%%%%%%%%%%%%%%%%%%%%%%%%%%%%%%%%%%%%%%%%%%%%%%%%%%%%
{ \section{Renormalizable Adjoint $SU(5)$}} 
%%%%%%%%%%%%%%%%%%%%%%%%%%%%%%%%%%%%%%%%%%%%%%%%%%%%%%%%%%%%%%%%%%%%%%%%%%%%%%%%%%%%%%%%%
In this section we will discuss the simplest renormalizable $SU(5)$ model 
where the neutrino masses are generated through Type I and Type III 
seesaw mechanisms. In our model~\cite{paper1} the Higgs sector is 
composed of ${\bf {{24}}_H}=(\Sigma_8,\Sigma_3,\Sigma_{(3,2)},\Sigma_{(\bar{3}, 2)},\Sigma_{24})$
$=({8},{1},0)\bigoplus({1},{3},$
$0) \bigoplus ({3},{2},-5/6)\bigoplus(\overline{{3}},{2},5/6)
\bigoplus ({1},{1},0)$, 
${\bf 45_H}=(\Phi_1, \Phi_2, $ $\Phi_3, \Phi_4, \Phi_5, \Phi_6, H_2)=({8},{2},1/2)\bigoplus$
$(\overline{{6}},{1},-1/3)\bigoplus ({3},{3}$
$,-1/3) \bigoplus(\overline{{3}},{2},-7/6)
\bigoplus({3},{1},-1/3) \bigoplus (\overline{{3}},{1},4/3) \bigoplus ({1},$

${2},1/2)$ and ${\bf {{5}}_H}=({1},{2},1/2)$ $\bigoplus({3},{1},-1/3)$. 
The field ${45}$ satisfies the following conditions: 
$({45})^{\alpha \beta}_{\delta} = - ({45})^{\beta \alpha}_{\delta}$, 
$\sum_{\alpha=1}^5 ({45})^{\alpha \beta}_{\alpha} = 0$, and 
$v_{45} = \langle 45 \rangle^{1 5}_{1}$ $= \langle 45 \rangle^{2 5}_{2}= \langle 45 \rangle^{3 5}_{3}$. 
Now, in this model the Yukawa potential for charged fermions reads as:
\begin{eqnarray}
V_{Y} &=& {10} \ \overline{{5}} \ \left( Y_1 \ {5}^*_H \ + \  Y_2 \ {45}^*_H \right) \ + \nonumber\\
&+& {10} \ {10} \ \left( Y_3 \ {5_H} \ + \ Y_4 \ 45_H \right) \ + \ h.c. 
\end{eqnarray}
and the masses for charged leptons and down quarks are given by:
\begin{eqnarray}
M_D &=& Y_1 \ v_5^* \ + \ 2 Y_2 \ v_{45}^* ,\\
M_E &=& Y_1^T \ v_5^* \ - 6 \ Y_2^T \ v_{45}^* , \label{GJ}
\end{eqnarray}
where $\langle{5}_H \rangle=v_5$. $Y_1$ and $Y_2$ 
are arbitrary $3 \times 3$ matrices. Notice that there are 
clearly enough parameters in the Yukawa sector to fit 
all charged fermions masses. See reference~\cite{potential} 
for the study of the scalar potential and~\cite{IPG} for the 
relation between the fermion masses at the high scale, 
which is in agreement with the experiment. 

The SM decomposition of the needed 
extra multiplet for type III seesaw is given by: 
$
{\bf 24}= (\rho_8,\rho_3, \rho_{(3,2)}$
$, \rho_{(\bar{3}, 2)},
\rho_{0})=({8},{1},0)\bigoplus({1},{3},0)$
$\bigoplus({3},{2},-5/6) \bigoplus(\overline{{3}},{2}$
$,5/6) \bigoplus({1},{1},0) $. 
In our notation $\rho_3$ and $\rho_0$ are the $SU(2)_L$ 
triplet responsible for type III seesaw and the singlet 
responsible for type I seesaw, respectively. 

The new relevant interactions for neutrino masses in this context 
are given by:

\begin{equation}
\label{neutpot}
V_\nu = c_i \ \overline{5}_i \ 24 \ 5_H \ + \ p_i \ \overline{5}_i \ 24 \ 45_H  \ + \ h.c. 
\end{equation}      
Notice from Eq.~(1) and Eq.~(4) the possibility to generate all fermion masses, 
including the neutrino masses, with only two Higgses : $5_H$ and $45_H$. 
Using Eq.~(\ref{neutpot}) the neutrino mass matrix reads as:

\begin{eqnarray}
M^\nu_{ij} & = & \frac{a_i a_j}{M_{\rho_3}} \ + \ \frac{b_i b_j}{M_{\rho_0}}, 
\end{eqnarray} 
where
\begin{equation}
a_i = c_i  v_5 \ - \  3 p_i v_{45}, 
\end{equation}
and
\begin{equation}
b_i  =  \frac{\sqrt{15}}{2} \left( \frac{c_i  v_5}{5} \ + \ p_i v_{45}  \right). 
\end{equation}   
The theory predicts one massless neutrino at the tree level. Therefore, we could 
have a normal neutrino mass hierarchy: $m_3=\sqrt{\Delta m_{sun}^2 + \Delta m_{atm}^2}$, 
$m_2=\sqrt{\Delta m_{sun}^2}$ and $m_1=0$,
or the inverted neutrino 
mass hierarchy: $m_3=0$, $m_2=\sqrt{\Delta m_{atm}^2}$ and 
$m_1=\sqrt{\Delta m_{atm}^2 - \Delta m_{sun}^2}$. $\Delta m_{sun}^2 \approx 8 \times 10^{-5}$ eV$^{2}$ 
and $\Delta m_{atm}^2 $
$\approx 2.5 \times 10^{-3}$ eV$^{2}$ are the mass-squared 
differences of solar and atmospheric neutrino oscillations, respectively. 

The masses of the fields responsible for the seesaw mechanisms are computed 
using the new interactions between ${\bf 24}$ and ${\bf 24_H}$ in this model:
\begin{equation}
V_{24} = m \ Tr (24^2) \ + \ \lambda \ Tr (24^2 24_H)  
\end{equation}
Once $24_H$ gets the v.e.v, $\langle 24_H \rangle = v \ diag (2,2,2, -3, -3)$ 
$/ \sqrt{30}$, the masses of the fields living in $24$ are given by: 
\begin{eqnarray}
M_{\rho_0} &=& m -\frac{\tilde{\lambda} M_{GUT}}{\sqrt{\alpha_{GUT}}},\\
M_{\rho_3} &=& m -\frac{3 \tilde{\lambda} M_{GUT}}{\sqrt{\alpha_{GUT}}},\\
M_{\rho_8} &=& m + \frac{2 \tilde{\lambda} M_{GUT}}{\sqrt{\alpha_{GUT}}},\\
M_{\rho_{(3,2)}} &=& M_{\rho_{(\bar{3},2)}} = m -\frac{\tilde{\lambda} M_{GUT}}{2 \sqrt{\alpha_{GUT}}},
\end{eqnarray}
where we have used the relations $M_V= v \sqrt{5 \pi \alpha_{GUT}/3}$, 
$\tilde{\lambda}= \lambda / {\sqrt{50 \pi}}$ and chose $M_V$ as the 
unification scale. Notice that when the fermionic triplet $\rho_3$ 
responsible for type III seesaw mechanism is very light, the rest 
of the fields living in ${\bf 24}$ have to be heavy if we do not assume 
a very small value for the $\lambda$ parameter.

Let us discuss the different contributions to proton decay. 
For a review on proton decay see~\cite{review}. In this model there 
are five multiplets that mediate proton decay. These are the 
superheavy gauge bosons $V=({3},{2},-5/6)\bigoplus(\overline{{3}},{2},5/6)$, 
the $SU(3)$ triplet $T$, $\Phi_3$, $\Phi_5$ and $\Phi_6$. The least
model dependent and the dominant proton decay
contributions in non-supersymmetric scenarios are mediated by the gauge bosons. 
Its strength is set by $M_V$ and $\alpha_{GUT}$. 
In order to satisfy the experimental lower bounds on 
proton decay we must have $M_V \geq (2 \times 10^{15}) \ 5 \times 10^{13}$ GeV if 
we do (not) neglect the fermion mixings~\cite{upper}.    
The different constraints coming from unification 
and proton decay issue have been studied in detail 
in reference~\cite{paper1}. \\

%%%%%%%%%%%%%%%%%%%%%%%%%%%%%%%%%%%%%%%%%%%%%%%%%%%%%%%%%%%%%%%%%%%%%%%%%%%%%%%%%%%%%%%%
\section{Supersymmetric Adjoint $SU(5)$} 
%%%%%%%%%%%%%%%%%%%%%%%%%%%%%%%%%%%%%%%%%%%%%%%%%%%%%%%%%%%%%%%%%%%%%%%%%%%%%%%%%%%%%%%%%
Let us discuss in this section the supersymmetric version of this model. 
In the minimal supersymmetric $SU(5)$~\cite{SUSYSU(5)} the 
MSSM chiral superfields are unified in $\mathbf{\hat{\bar 5}}$ and 
$\mathbf{\hat{10}}$, while its Higgs sector comprises $\mathbf{\hat 5_H}$, 
$\mathbf{\hat{\bar 5}_H}$, and $\mathbf{\hat{24}_H}$. Now, in order to write 
down the supersymmetric version of the realistic grand unified theory 
discussed above we have to introduce three extra chiral superfields, 
$\mathbf{\hat{45}_H}$, $\mathbf{\hat{\overline{45}}_H}$ and $\mathbf{\hat{24}}$. 
Therefore, our Higgs sector will be composed of $\mathbf{{\hat{\overline{5}}}_H}$, 
$\mathbf{\hat{{{5}}}_H}$, $\mathbf{\hat{24}_H}$, 
$\mathbf{\hat{45}_H} =( \hat{\Phi}_1, \hat{\Phi}_2, \hat{\Phi}_3, \hat{\Phi}_4, \hat{\Phi}_5, \hat{\Phi}_6, \hat{H}_2)$, 
and $\mathbf{\hat{\overline{45}}_H} =( \hat{\overline \Phi}_1, $
$\hat{\overline \Phi}_2, \hat{\overline \Phi}_3, \hat{\overline \Phi}_4, \hat{\overline \Phi}_5, \hat{\overline \Phi}_6, \hat{\overline H}_2)$. In this model the Yukawa superpotential for 
charged fermions reads as:

\begin{eqnarray}
{\cal{W}}_{0} &=& {\hat{10}} \ \hat{\overline{{5}}} \ \left( Y_1 \ \hat{\overline{5}}_H 
\ + \  Y_2 \ \hat{\overline{45}}_H \right) \ + \nonumber\\
&+& {\hat{10}} \ {\hat{10}} \ \left( Y_3 \ {\hat{5}_H} \ + \ Y_4 \ \hat{45}_H \right)  
\end{eqnarray}
where $Y_2$ is an arbitrary $3 \times 3$ matrix. As it is well-known 
the relation between the masses of $\tau$ lepton and $b$ quark, 
$m_b (M_{GUT})= m_{\tau} (M_{GUT})$, is in agreement with the 
experiment. Therefore, the $Y_2$ matrix must only modify the 
relation between the masses of quarks and leptons of the first and 
second generation.

Since in this section we are interested in the supersymmetric version 
of the model, a new matter chiral superfield has to be introduced 
only if we want to have the so-called matter parity as a symmetry 
of the theory. Matter parity is defined as 
$M=(-1)^{3(B-L)}=(-1)^{2S} R$, where $M=-1$ for all matter superfields 
and $M=1$ for the Higgses and gauge superfields. In the case 
that matter parity is not conserved the neutrino masses can be generated 
through the M-parity violating interactions $\epsilon_i \hat{\bar{5}}_i \hat{5}_H$ 
and $\eta_i \hat{\bar{5}}_i \hat{24}_H \hat{5}_H$. Particularly, in the second 
term we have an $SU(2)$ fermionic triplet needed for type III seesaw mechanism. 
However, we want to keep matter-parity as a symmetry of the theory 
to avoid the dimension four contributions to the decay of proton coming 
from $\lambda_{ijk} \hat{10}_i \hat{\bar{5}}_j \hat{\bar{5}}_k$ and have 
the lightest neutralino as a good candidate for the cold dark matter of 
the universe. 

The new superpotential relevant for neutrino masses in this context 
is given by:
\begin{equation}
{\cal W}_{1} = c_i \ \hat{\overline{5}}_i \ \hat{24} \ \hat{5}_H \ 
+ \ p_i \ \hat{\overline{5}}_i \ \hat{24} \ \hat{45}_H   
\end{equation}      
As in the non-supersymmetric model the Higgses in the $\mathbf{45}$ representation 
play a crucial role to generate masses for charged fermions 
and neutrinos as well.  

There are also new relevant interactions between $\mathbf{\hat{24}}$ 
and $\mathbf{\hat{24}_H}$ in this model:

\begin{eqnarray}
{\cal W}_{2} &=& m_{\Sigma} \ {Tr} \ \hat{24}_H^2 \ + \ 
{\lambda_{\Sigma}} \  {Tr} \ \hat{24}_H^3 \ + \  
m \ {Tr} \ \hat{24}^2 \nonumber \\
\ & + & \ \lambda \ {Tr} \ (\hat{24}^2 \hat{24}_H)  
\end{eqnarray}
Notice that there are only two extra terms since matter parity is conserved.
Our Higgs sector is composed of $\mathbf{\hat{5}_H}$, 
$\mathbf{\hat{\bar{5}}_H}$, $\mathbf{\hat{{45}}_H}$, 
$\mathbf{\hat{\overline{45}}_H}$ and $\mathbf{\hat{24}_H}$ and the 
additional interactions between the different Higgs 
chiral superfields in the theory are:

\begin{eqnarray}
{\cal W}_{3} &=& m_{H} \ \hat{\overline{5}}_H  \hat{5}_H \ + \ 
\lambda_H \ \hat{\overline{5}}_H  \hat{24}_H \hat{5}_H \nonumber \\
\ & + & \ c_H \ \hat{\overline{5}}_H  \hat{24}_H \hat{45}_H \ + \ 
b_H \ \hat{\overline{45}}_H  \hat{24}_H \hat{5}_H \nonumber \\    
\ & + & \  m_{45} \ \hat{\overline{45}}_H  \hat{45}_H \ + \ 
a_H \ \hat{\overline{45}}_H  \hat{45}_H \hat{24}_H 
\end{eqnarray} 

Notice the simplicity of the model. Unfortunately, 
the scalar sector of the non-supersymmetric grand unified 
theory proposed in reference~\cite{paper1} is not very 
simple since there are many possible interactions 
between ${\bf 5_H}$, ${\bf 24_H}$ and ${\bf 45_H}$. 

In this model there are several multiplets that 
mediate proton decay. We have the usual gauge $d=6$ contributions, 
the Higgs $d=6$ contributions, and the dimension five contributions. 
The most important proton decay contributions are 
mediated by the fields: $\tilde{T}$, $\tilde{\overline{T}}$, 
$\tilde{\Phi}_3$, $\tilde{\overline{\Phi}}_3$, $\tilde{\Phi}_5$, 
$\tilde{\overline{\Phi}}_5$, $\tilde{\Phi}_6$, and 
$\tilde{\overline{\Phi}}_6$. Let us discuss the different 
LLLL and RRRR contributions. The so-called LLLL 
effective operators, $\hat{Q} \ \hat{Q} \ \hat{Q} \ \hat{L}$, 
are generated once we integrate out the fields 
$\tilde{T}$, $\tilde{\overline{T}}$, $\tilde{\Phi}_3$, $\tilde{\overline{\Phi}}_3$, 
$\tilde{\Phi}_5$, and $\tilde{\overline{\Phi}}_5$. 
The RRRR contributions, $\hat{U}^C \ \hat{E}^C \ \hat{U}^C \ \hat{D}^C$, 
are due to the presence of the fields $\tilde{T}$, $\tilde{\overline{T}}$,   
$\tilde{\Phi}_5$, $\tilde{\overline{\Phi}}_5$, $\tilde{\Phi}_6$, and 
$\tilde{\overline{\Phi}}_6$. In reference~\cite{paper2} we have 
discussed how to suppress those contributions in order to satisfy 
the proton decay bounds. See also reference~\cite{Nath07} for 
an alternative way to suppress nucleon decay in this scenario.

%%%%%%%%%%%%%%%%%%%%%%%%%%%%%%%%%%%%%%%
\section{Summary and Outlook}
%%%%%%%%%%%%%%%%%%%%%%%%%%%%%%%%%%%%%%%%
In this talk I have discussed two new renormalizable 
grand unified theories based on $SU(5)$; we refer to 
these models as Renormalizable Adjoint $SU(5)$~\cite{paper1} and 
Supersymmetric Adjoint $SU(5)$~\cite{paper2}. 
In both models it is possible to generate 
all fermion masses, including the neutrino masses, with 
the minimal number of Higgses. These theories predict 
one massless neutrino at the tree level and the leptogenesis 
mechanism can be realized. The neutrino masses are generated 
through the type I and type III seesaw mechanisms.
The contributions to proton decay have been discussed in detail.
The models presented in this talk can be considered as the 
simplest renormalizable (SUSY) grand unified theory 
based on the $SU(5)$ gauge symmetry.

%%%%%%%%%%%%%%%%%%%%%%%%%%%%%%%%%%%%%%%%%%%%%%%%%%%%%%%%%%%%%%%%%%%%%%%%%%%%%%%%%%%%%%

\end{document}